\documentclass[useAMS,usenatbib,usegraphicx]{mn2e}

\def\hunit {\, {\rm km \, s^{-1} \,Mpc^{-1} } } 

\def\zacc {z_{\rm acc}} 
\def\uacc {u_{\rm acc}} 
 
\def\ztp {z_{\rm tp}} 
\def\utp {u_{\rm tp}}

\def\Omm{\Omega_{\rm m}} 
\def\Omlam{\Omega_{\Lambda}} 
\def\Omk{\Omega_{\rm k}} 
 
\def\Omde{\Omega_{\rm DE}} 

\def\dme{\Delta\mu_{\rm E}} 
\def\yd{ y_{\rm D}} 

\def\simlt{\mathrel{\lower0.6ex\hbox{$\buildrel {\textstyle <} 
  \over {\scriptstyle \sim}$}}}
\def\simgt{\mathrel{\lower0.6ex\hbox{$\buildrel {\textstyle >}
 \over {\scriptstyle \sim}$}}}

\def\newtwo { }

\title[On the luminosity distance and $\zacc$]
{On the luminosity distance and the epoch of acceleration}

\author[Will Sutherland and Paul Rothnie] 
{Will Sutherland$^{1}$\thanks{E-mail: w.j.sutherland@qmul.ac.uk} 
 and Paul Rothnie$^{1}$
\\
$^{1}$School of Physics and Astronomy, Queen Mary University of London, 
  Mile End Road, London E1 4NS, UK.  
} 

\begin{document}

\date{MNRAS - accepted 2014 Nov 05. Received 2014 Nov 5; 
  in original form 2014 Jun 11}

\pagerange{\pageref{firstpage}--\pageref{lastpage}} \pubyear{2015}

\maketitle

\label{firstpage}

\begin{abstract}

 Standard cosmological models based on general relativity (GR) 
 with dark energy 
  predict that the Universe underwent
 a transition from decelerating to accelerating expansion at a 
 moderate redshift $\zacc \sim 0.7$. 
  Clearly, it is of great interest to directly
 measure this transition in a model-independent way, 
 without the assumption that GR is the correct theory of gravity.  
 We explore to what extent supernova (SN) luminosity distance measurements
 provide evidence for such a transition: 
 we show that, contrary to intuition,
   the well-known ``turnover'' in the SN distance residuals 
 $\Delta\mu$ relative to an empty (Milne) model does not give firm evidence for
  such a transition within the redshift range spanned by SN data. 
 The observed turnover in that diagram is predominantly due to 
   the negative curvature in the Milne model,
  {\em not} the deceleration predicted by $\Lambda$ cold dark matter
  and relatives.  
  We show that there are several advantages
  in plotting distance residuals against a flat, non-accelerating model 
  $(w = -1/3)$, and also remapping the $z-$axis
  to $u = \ln(1+z)$;  we outline a number of useful and intuitive 
  properties of this presentation.  
 We conclude that there are significant complementarities
 between SNe and baryon acoustic oscillations (BAOs): 
 SNe offer high precision at low redshifts and give
 good constraints on the net {\em amount} of acceleration since $z \sim 0.7$, 
 but are weak at constraining $\zacc$; 
  while radial BAO measurements are probably superior for placing 
 direct constraints on $\zacc$. 

\end{abstract}


\section{Introduction}
\label{sec:intro} 
 The $\Lambda$ cold dark matter ($\Lambda$CDM) model has become 
 well established as the standard model of cosmology, 
 due to its very impressive fit 
 to a variety of cosmological observations, including
 CMB anisotropy \citep{wmap9h,planck-xvi}, large-scale
 galaxy clustering including the baryon acoustic oscillation
 (BAO) feature \citep{boss13}, and the Hubble diagram for distant supernovae
 (SNe; \citealt{betoule14}).  In $\Lambda$CDM and close relatives, the 
 mass-energy content of the Universe
 underwent a transition from matter domination to dark 
 energy domination in the recent past at a redshift $z_{me} \sim 0.33$;
 the transition from decelerating to accelerating expansion, 
 hereafter $\zacc$, was somewhat
 earlier, at a redshift $\zacc \approx 0.67$. In $\Lambda$CDM,
  these are given by $1 + \zacc = \sqrt[3]{ 2 \Omega_\Lambda / \Omm }$ 
 and $1 + z_{me} = \sqrt[3]{\Omega_\Lambda / \Omm }$, so 
  $1 + \zacc = \sqrt[3]{2} (1 + z_{me})$.  
 We see later that the value of $\zacc$ is relatively insensitive 
 to dark energy properties, assuming standard GR and simple
 parametrizations of the dark energy equation of state. 
 
 The most direct evidence for recent accelerated expansion comes 
 from the many observations of distant SNe at $0.02 < z \simlt 1.5$; 
  the early SN results 
 in 1998 \citep{hiz98, scp99} began a rapid acceptance
  of dark energy, due also to previous indirect evidence from large-scale
  structure \citep{esm90}, 
  the cluster baryon fraction 
  \citep{white93} and the Hubble constant \citep{ferr96}.    
  Strong independent support came from observation of the 
  first CMB acoustic peak defining a near-flat 
 universe \citep{boom00, maxima00}, 
 combined with decisive evidence for a low value of $\Omm$ from 
  the 2dF Galaxy Redshift Survey \citep{peacock01, psp02}.    
 In the past decade there has been a rapid improvement
 in the precision of observations in all these areas (see references above),
 most recently from the {\em Planck}, Baryon Oscillation Spectroscopic 
 Survey (BOSS) and Supernova Legacy Survey (SNLS) projects. 
 Current joint constraints are impressively consistent 
  with $\Lambda$CDM with $\Omm \simeq 0.30$ 
 and $H_0 \simeq 68.3 \hunit$ \citep{boss13, betoule14}.     
  
 Many deductions in cosmology are based on 
  six, seven or eight-parameter fits of extended $\Lambda$CDM to 
 observational data, which generally show good consistency with
 the six-parameter model and place upper limits on the 
  additional parameters. 
  However, given our substantial ignorance of the
 nature of dark energy, it is clearly interesting to ask what
  we can deduce with fewer assumptions, e.g. keeping
 the cosmological principle while dropping the assumption of 
  standard gravity.   
 In particular, fitting models of GR with dark energy to the data 
   produces a reasonably sharp prediction for the value of $\zacc$; 
 however, if the apparent cosmic acceleration is due to another cause
 such as modified gravity \citep{cfps}, a giant local void \citep{celerier}
 or other, this may not necessarily hold; therefore, it is of considerable
 interest to see what constraints we can place on $\zacc$ {\em without} 
  assuming specific models. 

 It has been shown by e.g. \cite{shap-turn} that the SN
  brightness/redshift relation does provide
 evidence for accelerated expansion independent of GR; but  
  direct evidence for past deceleration is less secure. 
 A number of other authors have explored GR-independent constraints
 on the cosmic expansion history, {\newtwo dark energy evolution} 
  and/or $\zacc$;
 { \newtwo e.g. 
 \cite{ss06} provide a broad review mainly focused on 
 dark energy reconstruction; \citet{catt-viss} explore various
 distance definitions related to $z$ or $y = z/(1+z)$;  
 \cite{cl08} derived constraints on $\zacc$ from SNe 
 assuming simple parametrizations of deceleration parameter $q(z)$; 
 \cite{clark-zunck} provide a method for non-parametric
 reconstruction of $w(z)$ (mainly from future high-quality data); 
 \cite{mort-clark} provide non-parametric
  estimates of $H(z)$;  
 and \cite{ngb13} give a comparison of several methods
 for estimating $w(z)$ from SNe data. } 
 Our work is partly related to these, 
  but focusing more on the possibility of non-parametric 
 constraints specifically on $\zacc$; 
 where we overlap we are generally in agreement. 

 The plan of the paper is as follows: in Section~\ref{sec:dl} 
 we discuss the value of $\zacc$ and the SN Hubble diagram,
  and the cause of the downturn in the latter.  
 In Section~\ref{sec:flatna} we point out several advantages of 
  comparing SN residuals relative to 
  a flat non-accelerating model.  
 We discuss some future prospects in Section~\ref{sec:disc},
 and we summarize our conclusions in Section~\ref{sec:conc}. 
 Our default model is $\Lambda$CDM with $\Omm = 0.300$; $H_0$ generally
  cancels except where stated. 

\section{Relation between luminosity distances and $\zacc$} 
\label{sec:dl} 

\subsection{The expected value of $\zacc$} 
\label{sec:zacc} 
\begin{figure*} 
\includegraphics[angle=-90,width=12cm]{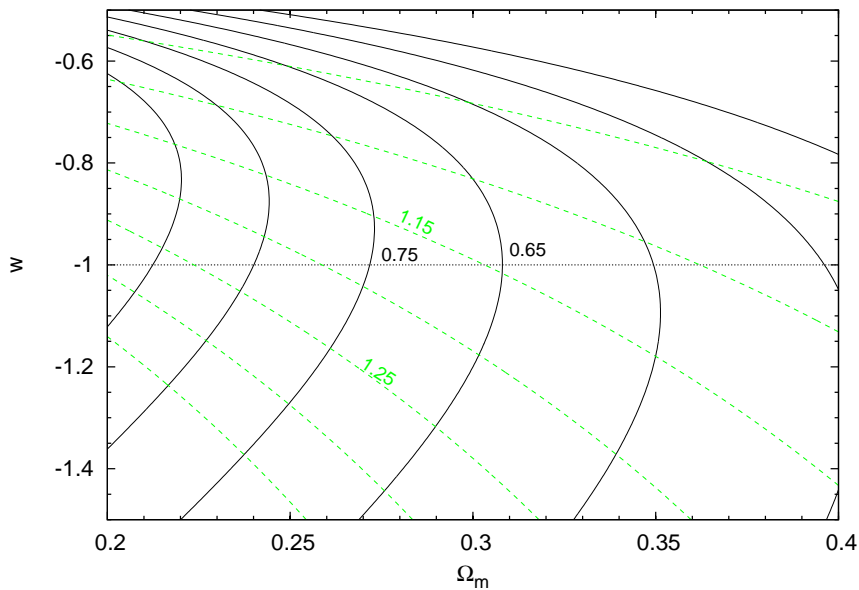} 
\caption{ 
 A contour plot of the acceleration redshift $\zacc$,
  and $(1+\zacc)/E(\zacc)$, as functions of
 $\Omm, w$  for flat wCDM models.  The dotted horizontal line 
  shows $w = -1$. \hfill\break  
  The solid black contours show $\zacc$, in linear steps of 0.1 from 
 0.35 (right) to  0.95 (left).  The dashed green contours show 
 $(1+\zacc)/E(\zacc)$ (i.e. total net speed-up) in linear steps 
  of 0.05 from 1.05 (upper right) to 1.35 (lower left).  Selected contours
  are labelled. 
\label{fig:zacc} 
}
\end{figure*}

 Here we note that the value \footnote{ In highly non-standard models, 
  it is not guaranteed that $\zacc$ (defined by $\ddot{a} = 0$) 
 is single-valued; 
 e.g. if there were short-period low-amplitude
  oscillations in $\dot{a}$, or a past accelerating
  phase transitioned back to deceleration at a very low redshift, 
  then in principle $\zacc$ may be multi-valued.
 These possibilities appear improbable and hard to test 
 observationally, 
  so we assume $\zacc$ is single-valued (after the CMB era) 
   for the remainder of this paper; see also \cite{linder10}.} 
 of $\zacc$ is now constrained rather
 well in flat $w$CDM models with constant dark energy 
 equation of state $w$; for this model family, $\zacc$ depends on
 only $\Omm$ and $w$, and is given by 
\begin{equation} 
\label{eq:zacc} 
  1 + \zacc = \left[ (-1 - 3w) (1-\Omm) / \Omm \right]^{-1/3w} \  
\end{equation} 
 (e.g. \citealt{tr02}). 
 This is shown in a contour plot in Fig.~\ref{fig:zacc}. 
 It is interesting that in the neighbourhood of $\Omm \sim 0.3, w \sim -1$,
  the contours of constant $\zacc$ are nearly vertical, 
 thus $\zacc$ is nearly independent of $w$ and is well 
  approximated by  
\begin{equation} 
  \zacc \simeq 0.671 - 2.65 (\Omega_m - 0.3) \ . 
\end{equation}  
 Qualitatively, this occurs because as $w$ increases above $-1$, 
  there is less negative pressure hence less acceleration
   per unit $\rho_{DE}$, 
  but larger $w$ gives higher $\rho_{DE}$ in the past;  these effects
  happen to cancel (largely coincidentally) near the concordance model, 
   so $\zacc$ is rather insensitive to $w$. 
 This has positive and negative consequences: on the one hand, measuring
  $\zacc$ is not useful for constraining $w$; on the other hand,  
 the range $0.60 \le \zacc \le 0.75$ appears to be a robust prediction of
  $w$CDM, so if future data 
  {\newtwo (e.g. direct measurements
  of $H(z)$ from BAOs or cosmic chronometers, 
  or new more precise SN data)}    
  were to empirically measure $\zacc$ {\em outside} this 
  range, it could essentially falsify the whole class of $w$CDM models.  
 (Models with time-varying $w$ such as the common model $w(a) = w_0 + w_a(1-a)$
  allow a wider range of $\zacc$, but these generally require $\zacc < 1$
  unless $w_a$ is dramatically negative, $w_a \simlt -1$, which is
  disfavoured in most quintessence-type models).  

 In Fig.~\ref{fig:zacc} we also show contours of $(1+\zacc)/E(\zacc)$,  
  which is equivalent to the ``net speedup'' or integrated acceleration
 between $\zacc$ and today; this is discussed later in \S~\ref{sec:flatna}.  

\subsection{SN data} 
\label{sec:sn} 

For comparison with models, we use the ``Union 2.1'' compilation of
  type-Ia SN distance moduli \citep{union21}, which contains
 580 SNe of good quality spanning the range 
 $0.01 < z < 1.6$.  For plotting purposes
 we divide the sample into bins of approximately equal width in
 $\ln(1+z)$, while adjusting bin widths so that each bin contains
  $\ge 20$ SNe except at the highest redshifts; then, the mean 
  distance modulus residual and weighted average redshift are computed
  for each bin. The resulting binned data points are 
   shown as `Union 2.1' in subsequent figures. 

\begin{figure} 
\hspace*{-5mm}\includegraphics[angle=-90,width=10cm]{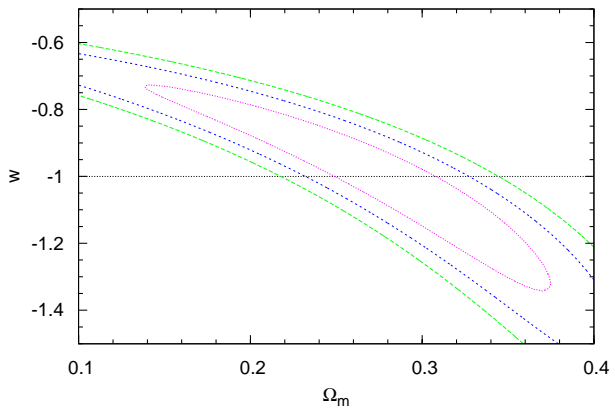} 
\caption{ 
 The allowed region in the $(\Omm, w)$ plane 
 from fitting flat constant-$w$ models to the Union 2.1 SN
 sample.  Contours show the values of $\Delta \chi^2 = 2.3, 6.0, 10.6$,
  corresponding to 68, 95 and 99.8 percent confidence regions. 
\label{fig:sn-omw} 
}
\end{figure}

We show a fit of this data set to flat $w$CDM models (with
 $\Omega_m$ and constant $w$ as the fit parameters; results of this
 fit are shown in Fig.~\ref{fig:sn-omw}, 
 with a best-fitting point near $\Omm = 0.28, w = -1.01$. 
    This shows the well-known
  degeneracy track between $\Omega_m$ and $w$; here we note that
 the long axis of the track is quite similar to the
  contour $(1+\zacc)/E(\zacc) \approx 1.15$ in Fig.~\ref{fig:zacc};
  this is discussed in later sections. 

We note that a more recent SN Ia compilation has
 been produced by \citet{betoule14} which includes
 more intermediate redshift SNe, 
  more detailed photometric calibration and expanded treatment of
 systematic errors; however, the best-fitting
 parameters from the latter paper are within  $1\sigma$
 of those above, so the slight difference is not important for
 the remainder of this paper.

\subsection{Fiducial models and $\Delta\mu$} 
\label{sec:dmu} 

 The observations of Type Ia SNe are sensitive to   
  the standard luminosity distance $D_L(z)$ for each SN,
 plus some scatter due to the intrinsic dispersion in absolute
  magnitude per SN.  In practice,  the distant $z \simgt 0.1$
 SNe are compared to a local sample ``in the Hubble flow'' 
 typically at $z \sim 0.02$ to $0.05$; for the local sample, 
 peculiar velocities are assumed to be relatively small compared 
 to the cosmological redshift, so the value of $H_0$ cancels
 with the (unknown) characteristic 
  luminosity $L_c$ of a standardized SN. Thus, quasi-local 
  SNe really constrain the degenerate combination $h^2 L_{c}$ 
 or equivalently $M_c + 5 \log_{10} h$; and comparison of distant
  and local SN samples actually constrains the distance ratio
  $D_L(z) / D_L(z \sim 0.03)$, rather than the absolute distance.  

 The value of $D_L(z)$ spans a very wide range over the redshift
 interval covered by SNe: from 
  $z \sim 0.03$ to $z \sim 2$ is a factor of $\approx 118$ 
  in distance or $10.3$ magnitudes,
 while the differences between models are relatively modest: 
  e.g. 15\,percent differences between
   $\Lambda$CDM and a zero$-\Lambda$ open model, 
  down to differences $\sim 2\,$percent between 
   $\Lambda$CDM and a $w = -0.9$ model. 
 This implies that plotting $D_L(z)$ versus $z$ directly is not very informative
  since model differences are very small compared to the plot range;  
  therefore it is common to present SN results 
 as residuals relative to some fiducial model; residuals are often
  presented in distance modulus or magnitude units, i.e. 
\begin{equation}
  \Delta\mu(z) \equiv 5 \log_{10} { D_L(z) \over D_{L,{\rm fid}}(z) }
\end{equation}  
  where $D_{L,{\rm fid}}$ is the value for some fiducial model.
 The choice of fiducial model is essentially arbitrary (up to small
  binning effects second-order in bin size); however, this choice can have
 a strong effect on the shape of the results and 
 intuitive deductions, as shown below. 
  
\begin{figure*} 
\includegraphics[angle=-90,width=15cm]{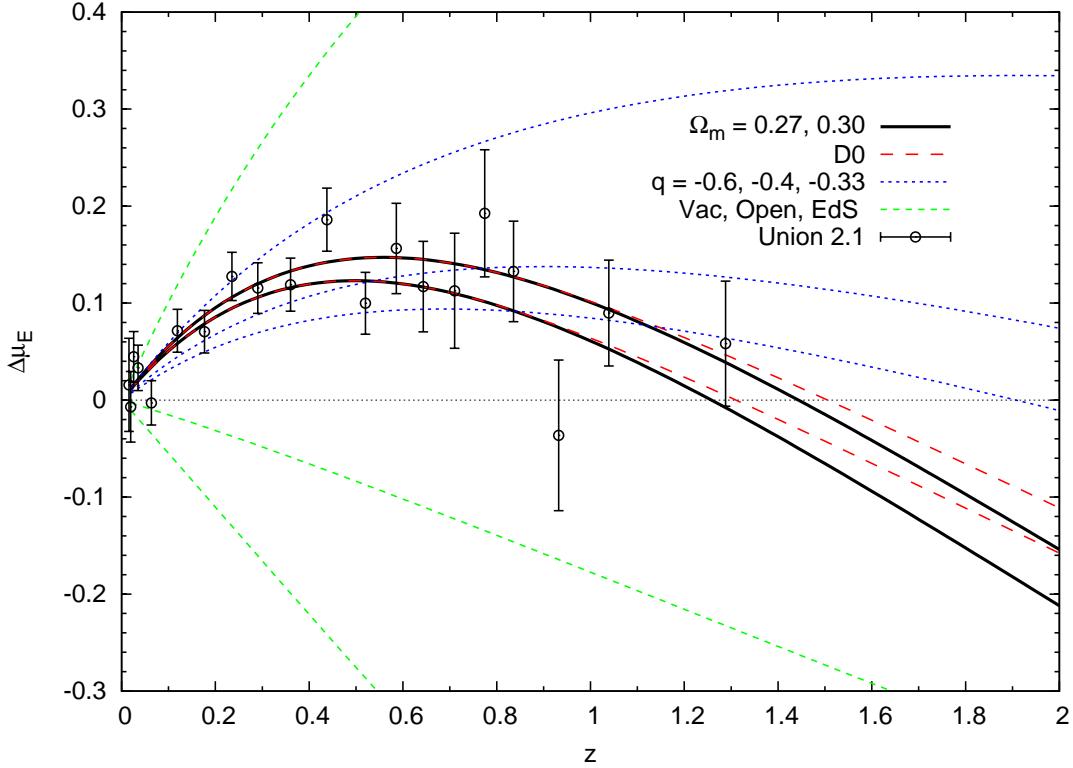} 
\caption{ 
 Distance modulus residuals relative to the Milne model for various
 cosmological models. The solid black lines show $\Lambda$CDM with 
  $\Omm = 0.27$ (upper) and $0.30$ (lower).  
 Long-dashed red lines show the corresponding D0 models (Equation~\ref{eq:d0}) 
  with deceleration artificially turned off above $\zacc$. 
  Dashed green lines show Friedmann models
 of historical interest: from top to bottom, 
 a pure-vacuum model ($\Omlam = 1$); an open model with $\Omm = 0.27$, 
  $\Omlam = 0$; and an Einstein-de Sitter model $(\Omm = 1)$.   
  Dotted blue lines show constant-$q$ models with
  $q_c = -0.6, -0.4, -0.33$ respectively (top to bottom). 
\label{fig:dmue} 
}
\end{figure*} 
 
 One obvious choice of fiducial is $\Lambda$CDM itself; however, this
 makes observed residuals (almost) flat--line, which does not translate readily 
  into inferences on deceleration or acceleration. 
 Another common choice of fiducial model is the empty or Milne model,
 with $\Omm = 0$, $\Omlam = 0$, $\Omk = 1$,
 as used by many notable papers e.g. \cite{hiz98, leib01, hiz04, gl11}.  
 The zero matter density means  
  this is clearly not a viable model for the real Universe, but 
  it is a convenient fiducial model for two reasons: 
 \begin{enumerate} 
 \item  It has a very simple analytic form for $D_L(z)$, given by 
 \begin{equation}
\label{eq:dle} 
  D_{L,E}(z) = \frac{c}{H_0} z \left(1 + \frac{z}{2}\right) \ ; 
\end{equation} 
 hereafter we define $\Delta \mu_E$ to be distance modulus residuals
  relative to this. 
\item For a given $H_0$, the Milne model has the maximum luminosity
 distance among all Friedmann models with zero 
  dark energy (assuming non-negative matter density).  Therefore,  
 observational evidence for distance ratios larger
 than the Milne model (positive $\Delta \mu_E$) at any redshift 
  is direct evidence that we do not live in a Friedmann 
 model with zero dark energy. 
\end{enumerate} 
{\newtwo However, using the Milne model 
 as fiducial has some drawbacks which we discuss
 in the next subsection; 
 we suggest an improved fiducial model in Section~\ref{sec:flatna} 
 (see also \citealt{mort-clark}). 
} 

\vfill
\newpage
\subsection{Downturn in distance residuals} 

 It is very well known that observed SN distance residuals 
  are all significantly positive at $0.2 \simlt z \simlt 0.6$, in agreement
 with the $\Lambda$CDM accelerating expansion. 
 It is also fairly well known that $\Lambda$CDM models exhibit 
 a turning point (a maximum) in the $\dme(z)$ relation. 
   Fig.~\ref{fig:dmue} shows that this turning point, hereafter $z_{tp}$,  
 occurs at $z \simeq 0.50$ for the $\Omm = 0.300$ concordance model, 
 and the predicted residuals then decline to a zero-crossing 
 at $z \simeq 1.26$. It is seen in Fig.~\ref{fig:dmue} that
 the actual supernova data do hint at the existence of a turnover, 
 with the three data points at $z > 0.9$ all slightly low compared
 to their predecessors.  
  The actual evidence for this turnover is not decisive, but it is
  clearly somewhat preferred by the data.  
 The turnover occurs quite close to the theoretical transition 
 epoch $\zacc \approx 0.67$, and it is therefore widely believed 
 (at least anecdotally) that supernovae have directly detected the 
  predicted cosmic  deceleration at $z \simgt 1$. 
 {\newtwo We discuss some prior claims to this effect in 
 Appendix~\ref{app:decel}. } 

 We demonstrate in the next subsection that the latter conclusion does not
  follow; specifically, while a downturn in $\dme$ {\em is} favoured by
  the data, the downturn predicted by $\Lambda$CDM is {\em mostly}  
 caused by the negative space curvature in the fiducial Milne model,  
 and cosmic deceleration makes only a 
  minority contribution to the downturn.  
 The fairly close match between $z_{tp}$ and $\zacc$ is found to be
 largely coincidental.  
 
\subsection{Cause of the turnover in $\dme$} 
\label{sec:turnover} 

 Assuming homogeneity, the luminosity distance $D_L(z)$ is given
  by 
\begin{equation}
\label{eq:dl} 
 D_L(z) = \frac{c}{H_0} (1+z) \, \frac{1}{\sqrt{\vert \Omk \vert} }  
   S_k\left( \sqrt{\vert \Omk \vert} \int_0^{z} \frac{ dz'}{E(z')} \right) 
\end{equation} 
 with $E(z) \equiv H(z)/H_0$, and 
 the function $S_k(x) = \sin x, x, \sinh x$ 
 for $k = +1, 0, -1$ respectively, where $k$ is the sign of the
 curvature (opposite to the sign of $\Omk$, in the usual convention
 where $\Omk = 1 - \Omega_{tot}$).  
  
 It is convenient to factorize this so that 
\begin{eqnarray} 
\label{eq:dldr} 
 D_L(z) & = & (1+z) \, D_R(z) \, \left( \frac{S_k(x)}{x} \right) \\ 
 D_R(z) & = &  \frac{c}{H_0} \int_0^z \frac{dz'}{E(z')}  \quad = 
  c \int_0^z \frac{dz'}{H(z')} \\
 x & \equiv & \frac{H_0 \, D_R(z) }{c} \, \sqrt{\vert \Omk \vert }  
\end{eqnarray} 
 where $D_R(z)$ is the comoving radial distance to redshift $z$; 
 and $x$ is the dimensionless ratio between $D_R(z)$ and the 
 cosmic curvature radius, which in a Friedmann model is
  $ R_c = c / H_0 \sqrt{\vert \Omk \vert } $. 
 We note that these distance results are still valid 
 in a homogeneous and isotropic non-GR model, 
 as long as the Robertson-Walker metric applies and 
 we define $\Omk$ from the curvature radius 
 via $\Omk \equiv \pm (H_0 R_c / c)^{-2}$,  
 which is then not necessarily equal to $1 - \Omega_{tot}$. 

 Looking at equation~(\ref{eq:dldr}), the first $(1+z)$ factor 
  is parameter-independent
  and due to time-dilation and loss of photon energy; 
 these each give one power of $(1+z)^{-1}$ in flux, hence combine
  to $(1+z)$ in equivalent distance.  
  The parameter dependence of $D_L(z)$ then factorizes into
  two parts, the $D_R(z)$ term dependent only on expansion history, 
 and the factor $S_k(x)/x$ which depends mainly on 
  curvature and also (more weakly) on expansion history;
  this is asymptotically $1 - k x^2/6$ for $x \ll 1$, or
 $1 + \Omega_k z^2 / 6$ for $z \ll 1$.  
 The factorization above is helpful to understand the relative importance 
 of curvature versus acceleration/deceleration on the distances and
 distance ratios.  In the non-flat $\Lambda$CDM model, the 
 combination of {\em Planck}+BAO  
 data requires $\vert \Omega_k \vert < 0.008$ at 95 percent
 confidence\footnote{ 
 We note that in non-GR models the standard 
  limits on $\Omega_k$ does not apply;
 however, if the true cosmology were a curved non-GR model, if 
 $\vert \Omega_k \vert \simgt 0.05$ we would then require a rather
  close cancellation between curvature and non-GR effects
  in order to make the non-flat $\Lambda$CDM fits turn out so 
  close to $\Omega_k = 0$.   
  If we discard this possibility as an unnatural conspiracy, 
  it is reasonable to assume 
  $\vert \Omega_k \vert < 0.05$, and in that case the curvature factor   
 $S_k(x)/x \approx 1 \pm 0.01$ for $z < 1.5$ for  
  reasonable expansion histories.}, 
 {\newtwo (see equations~68a and b of  \cite{planck-xvi})}, 
  which implies that 
  the curvature factor is within 0.2~percent of 1 at 
  the redshift range $z \simlt 1.5$ of current SNe. 
  
 It is now interesting to compare terms in equation~(\ref{eq:dldr}) 
  for the $\Lambda$CDM and empty models.  
 In the case of the empty model, $D_R(z) = (c/H_0) \ln(1+z)$, 
  $\Omega_k = +1$, so equation~(\ref{eq:dldr}) becomes 
\begin{equation} 
\label{eq:dlefac} 
 D_{L,E}(z) = (1+z) {c \over H_0} \ln(1+z) 
   \, { \sinh(\ln(1+z)) \over \ln(1+z)}  
\end{equation} 
  which easily simplifies to equation~(\ref{eq:dle}). 
 However, it is more informative to keep the longer form
  of equation~(\ref{eq:dlefac}) since 
  the rightmost fraction is a pure curvature effect; it is
 well approximated by $1 + (\ln(1+z))^2/6$ at $z \simlt 1$.    
   We next show that this term, {\em not} the transition
  to deceleration,  is the dominant cause of
  the downturn in $\Delta \mu_E$ for models similar to $\Lambda$CDM. 

 Considering the distance modulus residual $\Delta\mu$ for any flat
 model relative to the empty model, 
  we then have 
 \begin{eqnarray} 
\label{eq:dmue} 
 \Delta \mu_E(z) &=& 5 \log_{10} \left[ {\int_0^z {1 \over E(z')} dz' 
  \over \ln(1+z) } \,  {\ln(1+z) \over \sinh(\ln(1+z)) } \right] 
 \\
 \label{eq:dmusplit} 
   & \equiv &  \Delta \mu_{H}(z) - \Delta\mu_{k}(z) \\
\nonumber
\end{eqnarray}  
 where we have broken the $\Delta\mu_E$ into two additive terms, 
 $\Delta\mu_H (z) \equiv 5 \log_{10} [ \int_0^z (1/E(z')) dz' / 
  \ln(1+z) ]$ due to expansion histories, and 
 $\Delta\mu_{k} (z) \equiv 5 \log_{10} [\sinh(\ln(1+z))/\ln(1+z) ] $
  is the term due to curvature in the empty model (here defined so
 $\Delta\mu_k$ is positive,
  thus it is subtracted in equation~(\ref{eq:dmusplit}) above).  

 For illustration, we evaluate each of these terms for $\Lambda$CDM 
  (with $\Omm = 0.30$) at two specific redshifts: we
 choose $z_a = 0.50$ close to the turning point, 
 and $z_b = 1.26$ to be the downward zero-crossing where $\Delta\mu_E(z) = 0$. 
  We then find $\Delta\mu_E(0.50) = 0.1231 = 0.1822 - 0.0592$ where the
   latter two are $\Delta\mu_H$ and $\Delta\mu_k$ respectively.   
 At $z_b = 1.26$ we find $-0.0005 = 0.2350 - 0.2355$. Note that 
 $\Delta\mu_H$ grows from $z = 0.50$ to $z = 1.26$, since although
 the expansion is decelerating over most of this interval, 
 the expansion {\em rate} $\dot{a}$ remains smaller than the present-day 
  value; see below.    

 For comparison purposes, it is useful to evaluate
 how much the predicted deceleration contributes to $\Delta\mu_H$: for 
 this we define another model set, hereafter D0, 
  which exactly matches $\Lambda$CDM back to $\zacc$ but with deceleration
 artificially switched off $(q = 0)$ at $z > \zacc$: specifically,  
 we define model D0 by 
 \begin{eqnarray} 
\label{eq:d0}
   H(z) & =& H_{\Lambda CDM}(z)  \ \ {\rm if \ } z \le \zacc  \\ 
       & = &  H_{\Lambda CDM}(\zacc) (1+z) / (1+\zacc)  
  \ \ {\rm if \ } z > \zacc 
  \nonumber 
 \end{eqnarray} 
 The D0 models are somewhat artificial, but have a continuous
 $q(z)$ and are useful to isolate 
  the relative contribution of $\Lambda$CDM deceleration 
  on the observables. Also, they represent in a sense the closest 
 possible match to $\Lambda$CDM among all possible non-decelerating 
  models, {\newtwo so they are an interesting target to attempt to
 exclude observationally}. 
 The D0 model (for $\Omm = 0.30$) is identical to the 
  corresponding $\Lambda$CDM at $z_a$, 
  and at $z_b$ we find 
 $\Delta\mu_H = 0.2464$ (and $\Delta\mu_k = 0.2355$ again).  
 Therefore, the actual brightening effect attributable to deceleration in 
  $\Lambda$CDM is just the difference in $\Delta\mu_H$ between 
 $\Lambda$CDM and D0, which is only $-0.011$ mag.  This is smaller
 by a factor of 20 than the curvature effect; so, the bottom line
 of this subsection is that at $z = 1.26$,  95\% of this downturn 
 is due to curvature in the empty fiducial model (or 90\% if 
 we divide by the value $\Delta\mu_E = 0.1231$~mag at its maximum). Either
 way, it is clear that the open curvature in the Milne model 
  greatly dominates over deceleration as the source of the downturn 
 in $\Delta \mu_E$.

\section{An improved fiducial model} 
\label{sec:flatna} 
\subsection{The flat non-accelerating model} 
 We have argued above that the presentation of distance residuals
 from the Milne or empty model is potentially confusing, since
 it leads to a generic curvature-induced downturn in the residuals
 at $z \simgt 0.5$ which occurs {\em independent} of whether the
  expansion really decelerated prior to that epoch.  
 In this section we look at an improved fiducial model and
  demonstrate several advantages. 

 In particular, the above discussion suggests 
  a natural fiducial model is one with
  a constant expansion rate (deceleration
 parameter $q(z) = 0$, and $H(z) = H_0(1+z)$ 
 at all redshifts, as for the Milne model), but
 simply setting curvature to zero (equivalent to
  striking out the $\sinh$ in the equations above). 
   This is equivalent to
 a Friedmann model with $\Omm = 0$, $\Omega_{DE} = 1$ and
  $w = -1/3$; hereafter model N for short.   (This reference model has been 
  employed previously by \citet{seik-schw} and \citet{mort-clark},   
 but appears to be rather uncommon in the literature.) 
 Again, this model is not realistic due to the zero 
 matter density, but it is useful since it has both zero deceleration 
 and zero curvature. This model straightforwardly gives 
\begin{equation} 
 \label{eq:dln} 
 D_{L,N}(z) = \frac{c}{H_0} (1+z)\, \ln(1+z) \ . 
\end{equation} 
We now define the distance ratio for any other model, $\yd(z)$, as the ratio 
 $D_L(z)/D_{L,N}(z)$, therefore
\begin{equation} 
\label{eq:yd} 
  \yd(z) \equiv {H_0 D_R(z) \over c \ln(1+z)} \, {S_k(x) \over x}  
\end{equation} 
 For an almost-flat model at $z \simlt 1.7$ we can again  
  neglect the curvature term as very close to 1 (as per footnote 
 in Sect.~\ref{sec:turnover}  

Thus, for flat models the distance ratio becomes 
\begin{eqnarray}
 \label{eq:ydz} 
 \yd(z) &=& { 1 \over \ln(1+z) } \, \int_0^{z} \frac{dz'}{E(z')} 
\end{eqnarray} 
For many purposes below, it is more convenient to change the 
 redshift variable  to $u = \ln(1+z)$, which gives 
\begin{eqnarray} 
 \label{eq:ydu} 
  \yd(u) & = & \frac{1}{u} \, {\int_0^u \frac{1+z'}{E(z')} \; du' } \ ; 
\end{eqnarray} 
  as usual $u', z'$ are 
  dummy integration variables, not derivatives, and
 $\yd(u)$ means $\yd(z = e^u - 1)$.   
\begin{figure*} 
\includegraphics[angle=-90,width=15cm]{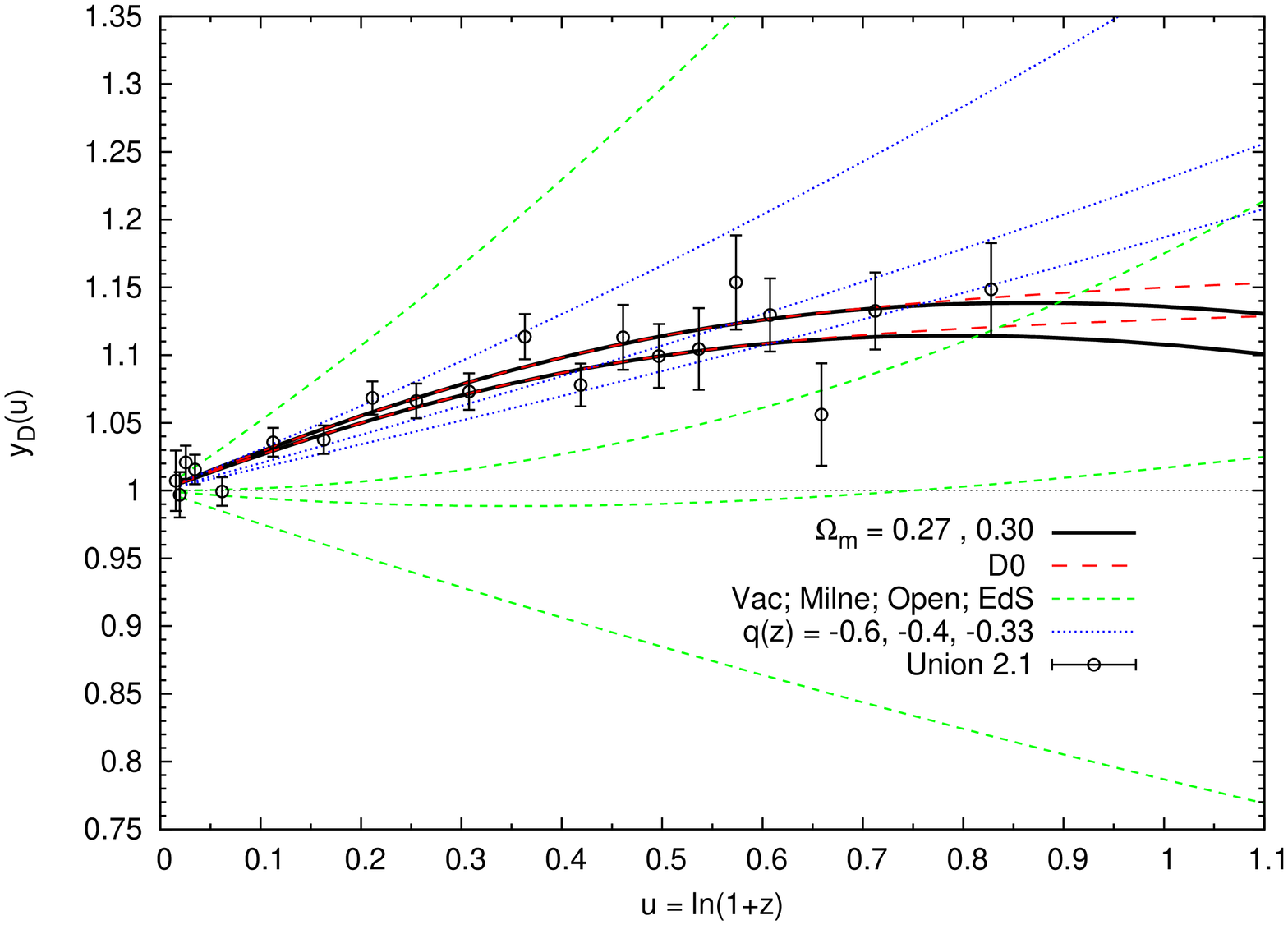} 
\caption{ As Fig.~\ref{fig:ydz}, but with the horizontal axis
  now linear in $u = \ln(1+z)$.  
\label{fig:ydu} } 

\includegraphics[angle=-90,width=15cm]{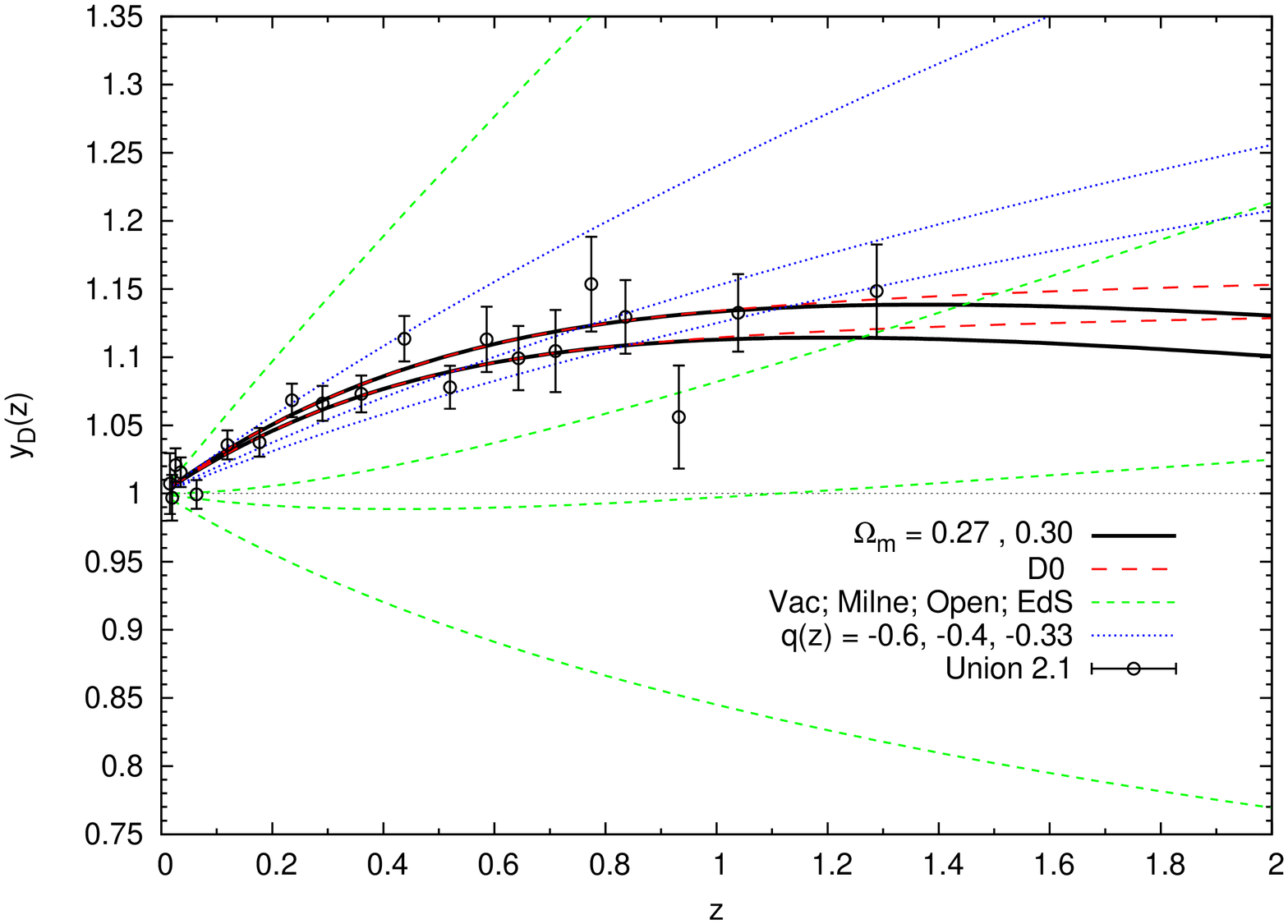} 
\caption{ The distance ratio $\yd(z)$ defined in equation~(\ref{eq:yd}) 
  for various cosmological models.  As in Fig.~\ref{fig:dmue}, 
 solid black lines are $\Lambda$CDM models with $\Omm = 0.27$ (upper)
  and $0.30$ (lower).  Long-dashed red lines are corresponding D0 models,
  with deceleration artificially switched off.  
 The short-dashed green lines are four Friedmann models of historical interest: 
 from top to bottom, 
  vacuum-dominated ($\Omm = 0, \Omlam = 1$);  empty (Milne); 
  open ($\Omm = 0.27, \Omde = 0$);  and Einstein-de Sitter ($\Omm = 1$).  
 Dotted blue lines are three 
 constant-$q$ models with $q = -0.6, -0.4, -0.33$ (top to bottom). 
  Points with errorbars show the binned Union 2.1 SNe data. 
\label{fig:ydz} }

\end{figure*} 

 This $\yd$ is directly related to $\Delta\mu_H$ above via 
  $\Delta\mu_H(z) = 5 \log_{10} \yd(z)$, 
  but several results below are simplified if we choose 
   {\em not} to apply this log.  
 Since $\yd(z)$ is fairly close to $1$ in reasonable models,
  this is anyway rather close to a linear stretch 
  $\Delta \mu_H \approx 2.17 (\yd -1)$. 

  Since $E(z)/(1+z)$ is just the
   expansion rate at $z$ relative to the present day, 
   i.e. $\dot{a}(z) / \dot{a}(z=0)$, 
  the integrand of equation~(\ref{eq:ydu}) is just the inverse of this; i.e.
  $\yd(z)$ measures the average value of $(\dot{a})_0 / \dot{a}$ 
  with respect to $\ln(1+z)$, over the interval from the source
 to the present.  It is more convenient to work with averages of $(1+z)/E(z)$ 
  rather than $1/E(z)$, since the former varies much more slowly
  with redshift:   
 for our default $\Lambda$CDM model,
   $(1+z)/E(z)$ reaches a maximum value of $1.153$ at $\zacc \simeq 0.67$, 
 crosses 1 again at $ z \simeq 2.08$, 
  and declines to 0.895 at $z = 3$. 

 Note also that since $(1+z)/E(z)$ contains the inverse of $\dot{a}$, 
 while $z$ increases backwards in time, derivatives 
 of $(1+z)/E(z)$ have the same sign as $\ddot{a}$, i.e. positive for
 acceleration.  
 In fact the standard deceleration parameter 
 $q \equiv   -\ddot{a}/(aH^2(a))$ is given by 
\begin{equation}
\label{eq:qu} 
 q(u) = - \frac{d}{du} \ln \left( {1+z \over E(z)} \right) 
\end{equation} 
 which is useful below. 

\subsection{Useful properties of $\yd$} 
\label{sec:ydprops} 

 The above definition of $\yd$ is simple and intuitive, 
 and we show below that it enables a number of useful 
 non-parametric deductions, as follows: 
 \begin{enumerate} 
 \item 
  It is clear above that a value of $\yd(z) > 1$ at any $z$ implies the
 past-average of $\dot{a}$ was less than the present value, i.e. 
 acceleration has dominated over deceleration over
 this interval (note, this is not strictly the same as requiring $\ddot{a} > 0$ 
  at the present day); this feature is similar to the Milne fiducial 
  model above. 
\item 
  It is easy to see that if $q(z)$ is always negative
  over some interval $0 \le z \le z_1$,  then $(1+z)/E(z)$ is a 
   strictly increasing function of $z$,
  and therefore so is $\yd(z)$; i.e. a flat model which is 
  non-decelerating at $0 < z < z_1$ 
  cannot have a turnover in $\yd$ at $z \le z_1$, regardless of the
  specific expansion history.  
 The converse of this is that {\em if} a turnover in $\yd(z)$ 
 is observed,   this implies a transition to deceleration must have occurred
  within the interval, i.e. we can definitely conclude
   $ \zacc < z_{tu} $ independent of the functional form of $E(z)$. 
  Also, if a turnover exists at $z_{tu}$, differentiating  
  equation~(\ref{eq:ydu}) implies that the value
  of $1/\dot{a}$ at $z_{tu}$ was equal to its
   average value (w.r.t. $u$) across the interval 
  from $z_{tu}$ to today. 
\item
 We can improve on the results above using the Mean Value Theorem:
 specifically, if had a known value $\yd(z_1) = y_1$, this theorem 
 implies that there exists some $z < z_1$ with $(1+z)/E(z) \ge y_1$; 
 i.e. the cosmic expansion rate has speeded up by at least a factor
  of $y_1$ since some $z < z_1$, independent of the functional 
 form of $E(z)$.    For a more realistic case where
 we measure an average value of $\yd$ in a finite bin, e.g. 
  $\langle \yd \rangle = \hat{y}$ averaged between $z_1 < z < z_2$, we can
 use the Mean Value Theorem {\em twice}: first, there exists some $z_m$
 within this bin with $\yd(z_m) = \hat{y}$, and secondly there exists
 some $z_3 \le z_m \le z_2$ satisfying $(1+z_3)/E(z_3) \ge \hat{y}$. 
 The above argument applies for exact knowledge of $\hat{y}$, neglecting
  error bars; however, it is clear that the same argument also applies 
  if we insert an observational lower bound for $\hat{y}$. 
\item
  Also, it is interesting to ask a reverse question:
  if the expansion was decelerating at all $z > \zacc$, does
 this imply that a turnover in $\yd(z)$ must exist ?  
 The answer appears to be `almost always': it is possible to
  build a contrived expansion history 
  where $q(z)$ crosses from negative to a small positive value, 
  then asymptotes back to zero from above at high $z$, so 
 $(1+z)/E(z)$ tends to a constant from above; in this 
   contrived case we can have deceleration at all $z > \zacc$ while
   $\yd(z)$ monotonically increases to the same constant.  
  However, if we assume non-infinitesimal deceleration, 
   $q(z) \ge +\epsilon$ for all $z > z_1$ and some positive 
   value $\epsilon$, it is readily proved that
  $\yd(z)$ must have a turnover at some $z$ 
 (though not necessarily in a readily observable range).  
\item 
  Differentiating equation~(\ref{eq:ydu}) and rearranging gives 
\begin{equation} 
 \label{eq:dyddu} 
 { 1 + z \over E(z) } = \yd(u) + u \frac{d\yd}{du} \ . 
\end{equation} 
  This gives us a direct graphical implication: taking the tangent to the 
   curve of $\yd(u)$ at any point $u_1$ 
  and extrapolating the tangent line to $u = 2u_1$ gives us directly  
  the value of $(1 + z) / E(z)$ at $z_1 = \exp(u_1)-1$. 
 
 Differentiating again shows that the transition to 
 acceleration occurs when $d^2\yd / du^2 = -(2/u) d\yd / du$; 
 however, as is well known the need to take a second derivative
 of noisy data  implies that this is not a very useful method
 for directly estimating $\uacc$. 
\item 
 Substituting from equation~(\ref{eq:qu}) above leads to the compact results 
 \begin{eqnarray}
\label{eq:ydq} 
  \yd(u) & = & \frac{1}{u} \int_0^u 
  \exp\left[ -\int_0^{u'} q(u'') \, du'' \right] \; du' \ , \\ 
 q(u) & = & { - 2 \frac{d\yd}{du} - u \frac{d^2 \yd}{du^2} \over 
   \yd(u) + u \frac{d\yd}{du} } \ ; 
\end{eqnarray} 
  this shows that $q_0 = -2 (d\yd/du)(0)$, but also that
 as $u$ increases we get increasing weight from the second-derivative
 term, so it becomes increasingly more challenging to constrain 
 $q(u)$ directly from numerical derivatives of data with realistic noise. 
 Even for optimistic $1\%$ error bars on $\yd$ in bins $\Delta u = 0.1$,
  we get order-unity errors on $d^2 \yd/du^2$, so 
 free-form reconstruction of $q(u)$ is essentially impossible 
 given realistic errors;  
 the best we can do is assume some smooth few-parameter model
  for $q(u)$ and fit.  
\item
 From equation~(\ref{eq:ydu}) it clearly follows that for two
  measurements at redshifts corresponding to $u_1, u_2$ we have 
\begin{equation} 
\label{eq:ydrange} 
{ u_2 \yd(u_2) - u_1 \yd(u_1) \over u_2 - u_1 } = 
  \frac{1}{u_2 - u_1} \int_{u_1}^{u_2} { 1 + z' \over E(z') } \, du' 
\end{equation}
 where the right-hand side (RHS) 
 is the average of $(1+z)/E(z)$ between the endpoints;   
 therefore we can estimate this average
   as a linear combination of the two values at the ends; 
 this is simple with respect to combination of error bars, 
 and does not assume $u_2 - u_1$ is small.   
\item
We now show another useful property of $\yd$: 
for any flat model with $q(z) =\ $constant (of either sign),  
the second derivative $d^2 \yd / du^2$ with respect to $u$ is everywhere
 non-negative.  
For such a model, denoting $q_c$ as the constant value of $q$, 
 we have $H(z) = H_0 (1+z)^{1+q_c}$. This easily leads to 
\begin{eqnarray} 
  D_L(z) &=& \frac{c}{H_0} (1+z) \frac{-1}{q_c} \left[(1+z)^{-q_c} -1 \right] \\ 
  \yd(z) &=& \frac{-1}{q_c} \frac{ (1+z)^{-q_c} -1 } {\ln(1+z)} \\ 
  \yd(u) &=& \frac{1 - e^{-q_c u}}{q_c u} 
\end{eqnarray} 
 Now differentiating twice with respect to $u$ gives 
 \begin{eqnarray} 
 \frac{d^2 \yd}{du^2} & = & \frac{-1}{q_c} \left[ \frac{ e^{-q_c u} 
   (u^2 q_c^2 + 2 u q_c +2) -2 }{u^3} \right] \\ 
   & = &  q_c^2 \left[ \frac{2 - e^{-p}(p^2 + 2p +2)}{p^3} \right]  
 \end{eqnarray} 
 where we define $p \equiv q_c u$. 
 The function in square brackets above is positive for all $p$, 
 thus the above second derivative is everywhere non-negative 
  for any value of $q_c$ with either sign, and is zero only if
  $q_c = 0$ and $\yd \equiv 1$.   
  For the cases of interest here,  we are mainly
  interested in $-0.6 < q_c < 0$ at $0 < u < 1$, hence
 $-0.6 < p < 0$; the square-bracket term 
  evaluates to $1/3$ for $p = 0$ and $0.53$ for $p = -0.6$,
  so for any reasonable $q_c$ model the second derivative
  is then between $0.33 q_c^2$ and $0.53 q_c^2$, i.e. small, positive
  and slowly varying with $u$. 

This has a useful consequence: if $q(u)$ were in fact any constant,  
 then the graph of $\yd(u)$ versus $u$ must always
 show positive curvature (concave from above). 
  Conversely, if the observed data points
 for $\yd(u)$ exhibit significant negative curvature over
 some interval, we can conclude
 that $q(u)$ increased with $u$ at some point within the observed
 interval, again regardless of the specific functional form.
 (Note this does not necessarily imply that $q(u)$ became positive, 
 merely that it increased with $u$ i.e. was less negative in the past.) 
\end{enumerate}  
 We note that in the above points, items (i)-(iv) apply
 whether we choose $z$ or $u$ as the redshift variable, but items
 (v)-(viii) only apply with $u$ as the variable; this suggests
 the latter is preferred. 

 For an illustration of the current data, we plot 
 $\yd(u)$ against $u = \ln(1+z)$ in Fig.~\ref{fig:ydu}. 
 Although this is a simple transformation of the $x-$axis from
 Fig.~\ref{fig:ydz}, the qualitative appearance is somewhat
 different due to the non-linear transformation, i.e.
 higher redshifts become squashed.  The apparent ``knee'' 
  in the $\Lambda$CDM models around $z \sim 0.5$ in Fig.~\ref{fig:ydz} 
  is significantly smoothed out with the $u-$axis, 
  and both $\Lambda$CDM models now look very close to simple parabolas
 (see below).  
  Also, the constant-$q$ models change curvature from 
  negative in Fig.~\ref{fig:ydz} to small and positive in
  Fig.~\ref{fig:ydu}, as derived above. 
  Comparing to the data, it is clear that
 the SNe data points do marginally prefer a negative curvature
  in $\yd(u)$,  but not overwhelmingly so. 

 To quantify this, we fit three models to the $\yd(u)$ data points: 
 a linear model, a quadratic, and the family of constant-$q$ models above; 
 we find that the quadratic model is preferred over the linear
 model by $\Delta\chi^2 =  3.5$ for 1 extra degree of freedom (d.o.f.), while
 the quadratic is preferred over the best constant-$q$ model by 
  $\Delta \chi^2 = 5.7$ for 1 extra d.o.f. This indicates
 that negative curvature in $\yd$ (increasing $q$) is preferred, but
  only at around the $2\sigma$ significance level.  We expand on the 
 quadratic model below.

\subsection{A quadratic fitting function for $(1+z)/E(z)$} 
\label{sec:quad} 

Here we note that it is interesting to consider a fitting function
 where $1/\dot{a}$ is a quadratic function of $u$, 
  specifically 
\begin{equation} 
\label{eq:quadu} 
  { 1 + z \over E(z) } = 1 + b_1 u - b_2 u^2 
 \end{equation} 
 with arbitrary constants $b_1, b_2$, and $u \equiv \ln(1+z)$ as before. 
 The minus sign above is chosen so that positive 
 $b_1, b_2$ leads to recent acceleration and
  past deceleration as anticipated, with $\uacc = b_1 / 2b_2$ from 
 equation~(\ref{eq:qu}).    
 This fitting function is not physically motivated, 
  but is useful since it provides a 
 very good approximation to models similar to $\Lambda$CDM 
 at $u < 1, (z < 1.72)$ (see Appendix~\ref{app:qjerk} for an approximate 
 explanation of this property), 
 and it gives several simple analytic results below.

  Fitting this function to the default
  $\Lambda$CDM $(1+z)/E(z)$ over $0 < u < 1$ $(z < 1.72)$ 
  gives best-fitting values $b_1 = 0.569$, $b_2 = 0.530$ with 
 an rms error of 0.28~percent, 
 and a worst-case error of $-0.8$~percent.
 (This fit becomes significantly worse above $z \simgt 2$, 
  and has a catastrophic zero-crossing at $u \sim 2$ ($z \sim 6.4$), 
  but it is good over the range accessible to medium-term SN data.)   
 The functional form (\ref{eq:quadu}) gives simple relations between
  $\uacc$ and the turnover in $\yd$; it easily gives  
 \begin{eqnarray}
\label{eq:ydquad} 
 \yd(u) & = & 1 + \frac{1}{2} b_1 u - \frac{1}{3} b_2 u^2 \ ; \\ 
\label{eq:qquad} 
 q(u) & = & { -b_1 + 2 b_2 u \over 1 + b_1 u - b_2 u^2 } \ ; \\ 
\label{eq:dlquad} 
  D_L(u) & = & {c \over H_0} (1+z)  
   (u + \frac{1}{2} b_1 u^2 - \frac{1}{3} b_2 u^3) \ ; 
\end{eqnarray} 
  so $\yd(u)$ is also an exact quadratic in this case. 
 The $q(u)$ behaviour is approximately linear at moderate $u$,
 so this model is fairly similar to the model $q(a) = q_0 + q_a(1-a)$ 
  used elsewhere.  
 Equation~(\ref{eq:dlquad}) with values $b_1, b_2$ as above 
  matches the exact numerical $D_L(z)$ for $\Lambda$CDM with 
 very high accuracy, a maximum error only 0.13~percent back to $u = 1$; 
 this error is substantially smaller than for $E(z)$, due to the integral for 
 $D_L$.   

 We find that the results above also work well for $w$CDM models in the region
 $0.2 < \Omm < 0.4, -1.2 < w < -0.8$; thus, it is interesting (and
 partly a coincidence) that any $w$CDM model within the 
 presently-favoured range leads to a $\yd(u)$ curve virtually 
  indistinguishable from a quadratic, to around the 0.2 percent 
  level i.e. comparable to the line thickness in Fig.~\ref{fig:ydu}. 
 This gives another helpful feature: any proof of `percent-level'
  deviation of $\yd(u)$ from a simple quadratic would signify a failure of
  $w$CDM. 

 We now look at the relation between $\zacc$ and the turning point in $\yd$. 
 In the above model equation~(\ref{eq:quadu}) with $b_1,b_2 > 0$, recall
 the acceleration epoch is $u_{acc} = b_1 / 2 b_2$, hence 
  $(1+\zacc)/E(\zacc) = 1 + b_1^2 / 4b_2$ ; while the maximum
  in $\yd$ occurs at $u_{tp} = 3 b_1/ 4 b_2$, at height 
  $\yd(\utp) = 1 + 3 b_1^2 / 16 b_2$. 
 So, in this model $\zacc$ is  
  directly related to the location $\utp$ of the maximum, 
 and $(1+\zacc)/E(\zacc)$ is directly related to its height, via
\begin{eqnarray} 
\label{eq:zaztp} 
 \uacc & = & \frac{2}{3} \utp \ , \qquad  
  \zacc =  (1+\ztp)^{2/3} - 1 \ ;  \\ 
  { 1+\zacc \over E(\zacc)}  & = &  1 + \frac{4}{3} (\yd(\utp ) - 1 )  
\label{eq:ezacc} 
\end{eqnarray}
 without requiring to solve for $b_1, b_2$.   

\begin{figure} 
\hspace{-1.5cm}\includegraphics[angle=-90,width=11cm]{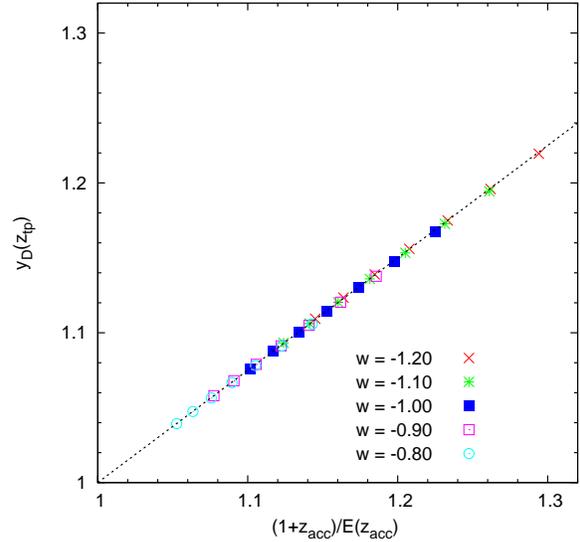} 
\caption{ 
 This figure shows 
 the peak value of $\yd$ against the integrated acceleration 
 $(1+\zacc)/E(\zacc)$, for a grid of $w$CDM models.  The differing 
 point styles show $w = -1.2, -1.1, -1.0, -0.9, -0.8$ as indicated 
  in the key. For each value of $w$ we show seven points with 
 $\Omm = 0.24, 0.26, 
  \ldots, 0.36$ in linear steps of $0.02$; 
  in each case these run from $\Omm = 0.24$ 
  at upper-right to $0.36$ at lower-left, 
   so the central point is $\Omm = 0.30$. 
 The dotted line (not a fit) is equation~(\ref{eq:ezacc}).  
\label{fig:ezacc} 
}
\end{figure}

 This suggests that for other reasonably smooth parametrizations of $E(z)$ 
  such as $w$CDM models, 
 we may expect equations~(\ref{eq:zaztp}) and (\ref{eq:ezacc}) 
  to hold approximately, rather 
 than exactly as above. 
 {\newtwo In our default $\Lambda$CDM model,  
  the exact values are $\zacc = 0.671$, $ (1+\zacc)/E(\zacc) = 1.1530$, 
  while from numerical evaluation 
  of $\utp$ and $\yd(\utp)$ the RHS of the above equations 
  evaluate to 0.693 and 1.1525 respectively; 
 thus equation~(\ref{eq:zaztp}) is quite good, while 
 equation~(\ref{eq:ezacc}) is an excellent approximation.     
  More generally, we have tested these for 
   wCDM models (constant $w$) with the results shown in 
 Fig.~\ref{fig:ezacc} ; this shows that equation~(\ref{eq:ezacc}) 
 remains very accurate for a substantial 
  range around the concordance model. } 

 We have also tested linear-$q$ models 
 $q(a) = q_0 + q_a(1-a)$, 
  and find that equation~(\ref{eq:ezacc}) is accurate to 
  better than 0.01 for reasonable values of $q_0, q_a$, 
  while equation~(\ref{eq:zaztp}) is somewhat worse but generally good 
  to a few percent.  
  For varying-$w$ models of the form $w(a) = w_0 + w_a(1-a)$, 
  these approximations remain good for $w_a \ge 0$ but 
 become somewhat less accurate for negative $w_a$, especially for 
 $w_a < -0.5$.  

The summary here is that equation~(\ref{eq:ezacc}) is generally an
 excellent approximation for constant-$w$ models, and a good  
 approximation for varying-$w$ if $w_a$ is not too negative; while 
 equation~(\ref{eq:zaztp}) is fairly good at the few-percent level.  

{\newtwo These approximations are useful since the right-hand-side
 of equations~(\ref{eq:zaztp}) and (\ref{eq:ezacc}) are in 
 principle directly observable: }  
 it is clear from Fig.~\ref{fig:ydu} 
 that the location of the possible maximum in $\yd$ is relatively 
 poorly constrained, but {\em if} the suggestion of negative curvature
 in $\yd$ is real and persists as expected to higher redshifts, 
 then the SNe datapoints imply that 
 $\yd(u)$ is probably approaching a maximum value $\sim 1.10 - 1.14$
  at $\utp \simlt 1$; if so, 
 this would give a direct and reasonably model-independent 
  inference of the integrated acceleration  
 $(1+\zacc)/E(\zacc) \approx 1.13 - 1.18$.  
 {\newtwo This provides a useful intuitive explanation of the 
 ridge-line of $\Omm$ versus $w$ observed in Fig.~\ref{fig:sn-omw}. 
 } 

 To summarize this subsection, we find that $w$CDM models with 
 constant $w$ near the concordance model 
  are very well approximated by the above fitting functions,
 i.e. very close to simple quadratics in $\yd(u)$, and thus 
 equations~(\ref{eq:zaztp}) and (\ref{eq:ezacc}) provide
  quite accurate approximations relating the observable turning point 
  in $\yd$ to $\zacc$ and the net acceleration. 

 Finally, in Appendix~\ref{app:dlapp} we use the fitting function 
 of equation~(\ref{eq:quadu}) to provide 
 a simple and accurate `computer-free' approximation to the luminosity 
 distance in $w$CDM models.

\subsection{Linear q(a) models} 
\label{sec:linqa} 
\begin{figure} 
\hspace{-1cm} 
\includegraphics[angle=-90,width=11cm]{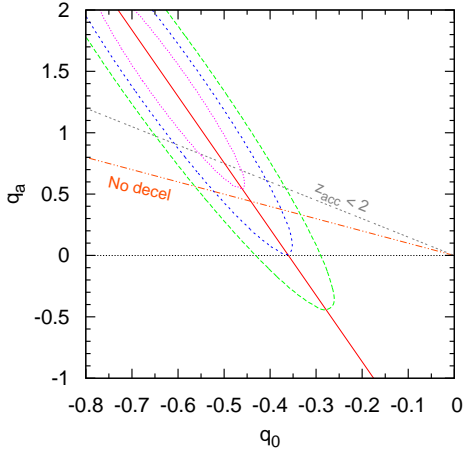} 
\caption{ 
 The allowed region in the $(q_0, q_a)$ plane 
 from fitting models with $q(a) = q_0 + q_a(1-a)$ 
 to the Union 2.1 supernova data.  
  Elliptical contours show the values of $\Delta \chi^2 = 2.3, 6.0, 10.6$
  corresponding to 68, 95 and 99.8 percent confidence regions. 
 The sloping lines bound the region of no deceleration and the region
 $\zacc < 2$, with the wedge between these giving $\zacc > 2$. 
 The line along the major axis of the ellipse is illustrative and
  gives a pivot value  
  $q(a = 0.815) = -0.36$ at $a = 0.815$ ($z = 0.227$). 
\label{fig:q0qa} 
}
\end{figure}

Here we briefly consider the two-parameter model family with
 deceleration parameter $q$ given by a linear function of scale factor
 $a$, i.e. 
\begin{equation} 
\label{eq:qa} 
 q(a) = q_0 + q_a(1-a) 
\end{equation} 
 for constants $q_0, q_a$. This model has been used before
 by various authors (e.g. \citealt{cl08}, \citealt{sca11}), 
  since it is simple, fairly flexible and can produce
 a fairly good approximation to the behaviour of 
  many dark energy models at $z \simlt 2$. 
 We have fitted this parameter pair to the Union~2.1 SN data, 
 with best-fit values at $(-0.62, +1.40)$ and
 the resulting likelihood contours shown in Fig.~\ref{fig:q0qa}; 
  as expected, negative $q_0$ is required at very high significance. 
  (This agrees well with a similar figure in \citet{sca11}). 
 The figure also shows  lines bounding the regions of no past
  deceleration $q_0 + q_a < 0$, 
 and the region $\zacc < 2$ equivalent to $q_0 + 2q_a / 3 > 0$; 
  the wedge between these lines corresponds to
  a transition redshift $\zacc > 2$.  
 This plot shows that the no-deceleration
  region is disfavoured at around the $1.3\sigma$ confidence level, 
 but there is a region inside the wedge $\zacc > 2$ which is allowed 
  at around $0.8\sigma$.  In this wedge, no deceleration occurs
   within the redshift range of observed SNe, 
  so the inference of deceleration relies 
  on a linear extrapolation of the $q(a)$ model beyond the range of 
  SNe.   
 This generally agrees with our previous conclusions, that
 a trend of less negative $q$ at higher redshift is clearly preferred,
 but there is negligible evidence from SN data alone 
 for an actual transition to deceleration within the observed range. 

\section{Discussion} 
\label{sec:disc} 
 It is instructive to blink back and forth between Figs~\ref{fig:dmue},
 \ref{fig:ydz}, \ref{fig:ydu} above: although from a parameter-fitting
 perspective there is no difference since the residuals
  (data--model) are all the same, from the perspective of visual
  intuition about expansion rate there are rather striking differences
 between these three Figures.  
  Clearly, Fig.~\ref{fig:dmue}
 shows a fairly convincing turnover in the data points; while 
  in Fig.~\ref{fig:ydz} the data shows negligible evidence for a turnover,
  but a reasonably convincing change in slope to a broad near-flat ``plateau'' 
  above $z \simgt 0.6$.  Finally, in 
 Fig.~\ref{fig:ydu} the $\Lambda$CDM models are extremely close to 
  parabolic (i.e. near-constant negative second derivative), 
 while the data points show near-linear behaviour with 
 a reasonable but non-decisive indication of negative curvature;   
 the constant-$q$ models show weak positive
 curvature as derived earlier in equation~(\ref{eq:dyddu}). 
 As we argued earlier, the turnover in Fig.~\ref{fig:dmue} is largely
 attributable to the negative space curvature in the Milne model, 
 not due to actual deceleration.  Figs~\ref{fig:ydz} and \ref{fig:ydu}
 show a much more gradual turnover in the $\Lambda$CDM models, 
 while the D0 models show the expected gradual rise; 
 clearly the current data are completely unable to discriminate
   between $\Lambda$CDM  and D0 models.  
  We suggest that Fig.~\ref{fig:ydu} is the most 
  informative due to the various useful intuitive 
 properties outlined in \S~\ref{sec:ydprops} above.

 The above conclusions seem somewhat unexpected: there
 is a widespread view {\newtwo (see Appendix~\ref{app:decel}) }   
 that the SN data has convincingly 
 verified the expected deceleration of the universe 
  at $z \simgt 1$.  However from the discussion above, 
  the SNe data are almost entirely inconclusive 
  on the sign of $q$ at $z > 0.7$, and even
  a constant-$q$ model with $q(z) \approx -0.4$ back to 
  $z \simgt 1$ is only excluded at the $\sim 2.5\sigma$ level which is
   significant but not overwhelming. 
  Thus, there is moderately good evidence for $q$ increasing in the past,
  but concluding that $q$ actually crossed zero 
  to a positive value relies strongly 
  on a smooth extrapolation of this trend, and 
   is therefore model-dependent.  

  Conversely, {\em if} we assume GR, almost all the acceptable
   models imply significant deceleration at $z > 1$. 
  Essentially, {\em if} we assume GR with the weak energy
  condition and a value of $\Omega_m > 0.2$, then the eightfold increase
  in $\rho_m$ back to $z=1$ combined with the much slower increase in
  dark energy guarantees matter domination and  
  deceleration at $z > 1$; in this case deceleration at $z > 1$ 
  is mainly a prediction of GR, rather than a feature 
   directly required by data. 
  For the value of $\zacc$ it is important
  to keep clear the distinction between an extrapolation based
 on GR parameter-fitting, or an actual detection 
  purely based on data.  

 It is clear that the CMB does provide much stronger constraints 
 due to the long distance lever-arm: if we assume the standard
 sound horizon length inferred from {\em Planck}, then we deduce 
  $\yd(z \simeq 1090) \simeq 0.44$, which clearly requires a turnover 
  and hence deceleration. 
  However, since
  the CMB only gives us one integrated distance to $z \sim 1090$
   spanning seven $e-$folds of expansion,
 while the supernova data constrains only the last one $e-$fold of expansion,  
 it would be straightforward to construct `designer' expansion histories
 with some extra deceleration hidden in the un-observed six $e-$folds to
 offset an absence of deceleration back to $z \sim 1.7$. 
 This is clearly contrived, 
  but would not directly conflict with any available $D_L(z)$ data. 
   Therefore, even adopting the standard distance constraint from 
  the CMB, we do not yet have a GR-independent proof 
  that the expansion was actually decelerating at $1 \simlt z \simlt 2$; 
  this is clearly the most probable and least contrived interpretation, 
  but loopholes remain.  
   
 We note that recent BAO results do 
  provide significant evidence for deceleration; from the first detection 
 of BAOs in the Ly-$\alpha$ forest by \citet{busca13} and comparison
 with lower--redshift measurements, \citet{busca13} quote  
\begin{equation} 
  { E(z=2.3)/3.3 \over E(z=0.5)/1.5 } = 1.17 \pm 0.05 
\end{equation}   
  which is a $3.4\,\sigma$ detection of deceleration between 
 the above two redshifts (though this 
  does assume an external WMAP7 curvature constraint, 
   which introduces some slight level of GR-dependence).  
However, the desirable goal of verifying that $\zacc < 1$ as expected
 is considerably more challenging, since the 
 expected change in $\dot{a}$ between $z = 0.67$ and $1$ 
  is only 1.7 percent in our default model. 
   The {\em Euclid} spacecraft \citep{euclid-red} is predicted 
 to get sub--percent measurements of $r_s H(z)$ at a range
 of redshifts $0.9 < z < 1.8$, which looks very promising for a direct
 model-independent result, {\newtwo 
 while improved ground-based measurements spanning
  $0.3 < z < 0.9$ would also be highly desirable.}

\section{Conclusions} 
\label{sec:conc} 
 We summarize our conclusions as follows: 
\begin{enumerate} 
\item The {\em predicted} value of $\zacc$ is rather well constrained 
  by current data within $w$CDM models, and is mainly sensitive
 to $\Omm$ rather than $w$;  
  this implies that a direct measurement of $\zacc$ is not helpful
 for measuring $w$, but is potentially an interesting 
  test of $w$CDM versus alternate models such as modified gravity. 
\item Contrary to intuition, the (probable) downturn in SN 
  residuals relative to the empty Milne model does {\em not} provide
 convincing evidence for deceleration. The predicted downturn is
 strongly dominated by the negative space curvature in the Milne model, and
 the actual deceleration in $\Lambda$CDM makes only a small
  minority contribution to the downturn.  
\item There are many advantages to presenting SNe distance 
  residuals relative
 to a flat coasting model ($\Omm = 0$, $\Omde = 1$, $w = -1/3$),
 and also in changing the horizontal axis from $z$ to $u = \ln(1+z)$ 
 as in Fig.~\ref{fig:ydu}.   
 This presentation enables a number of robust non-parametric deductions
  about expansion history based on the {\em global shape} of the
   observed residuals $\yd(u)$, without needing specific 
  numerical derivatives of data or fitting functions.   Notably,
 a turnover in this plot is decisive evidence for deceleration, 
  while any negative curvature in the data points is evidence for 
   higher $q$ in the past. 
\item If a turning point in $\yd(u)$ is observed, then we can 
  infer $\zacc$ from its location and $(1+\zacc)/E(\zacc)$ from its
 height from Eqs.~(\ref{eq:zaztp},\ref{eq:ezacc}); 
  the latter relation holds to very good accuracy in the case of $w$CDM models, 
  slightly degrading in the case of large negative $w_a$. 
\item For the case of $w$CDM models near the concordance range, 
  the model curves of $\yd(u)$ are remarkably close to simple quadratics 
  to an rms accuracy $\simlt 0.3$ percent, significantly better than 
  present data. This provides a simple intuitive visual test for 
  potential deviations from $w$CDM.  
\item For constraining expansion history, there are
  significant complementarities between SNe and BAO (or cosmic chronometers): 
  the SNe have a precise local anchor at $z \le 0.05$ and therefore
  place strong constraints on the {\em integrated} 
   acceleration, e.g. giving robust lower bounds on 
  the value of $1.7/E(0.7) \ge 1.1$.   
 However, the combination of the integral in
 SNe distances and the broad maximum in $(1+z)/E(z)$ around
 the acceleration transition implies that SNe are weak at giving
 model-independent constraints on $\zacc$.   
 In contrast, BAOs offer direct access to $H(z)$ without differentiation 
  and are therefore potentially stronger at constraining $\zacc$;
  but they have limited precision due to cosmic variance at
 $z \simlt 0.25$, and they are therefore weaker at constraining 
  the total integrated acceleration, most of which occurs 
  at $0 < z < 0.3$. 

 It is clearly important to get a good cross-anchor between SN
  measurements and BAO measurements for constraining the absolute
 distance scale; as argued by e.g. \cite{suth12}, precision measurements
  of {\em both} SNe and BAO at matched redshifts would be very useful for 
  this; {\newtwo see also \cite{blake11} for a slightly different but related
  approach}.   
\end{enumerate}

\section*{Acknowledgements} 
We thank the anonymous referee for helpful comments which have
 improved the clarity of this paper. 

 This is a pre-copyedited, author-produced version of
 an article accepted for publication in MNRAS. The version of
 record is available online, at Digital Object Identifier  
   DOI:10.1093/mnras/stu2369 \ . 

\vfill


 

\vspace{5mm}   

\appendix
\section[]{Previous claims of deceleration} 
\label{app:decel} 
{\newtwo Here, we provide a short discussion of previous claims 
 concerning evidence of past deceleration from SN data;
 these are mostly in press releases or the semi-popular literature,  
 but have had a significant influence. One of the earliest such claims 
  appears to be a quote from A.~Riess on NBC News\footnote{
 http://nbcnews.com/id/3077854 } 
 , 2001 April 02, 
 {`the new supernova, dubbed SN 1997ff, confirms that the universe
  began speeding up relatively recently'}. 
 A notable Scientific American article (Feb. 2004) by Riess \& Turner
  includes the quote `{the observations} (six SNe $>$ 7 Gyr old)
 { confirmed the existence of an early slowdown period.'} 
  This appears to be partly based on \cite{tr02}, and  
  that paper fits two classes of model: first $w$CDM models
 (in which case deceleration occurs almost `by assumption' 
   for reasonable values of $\Omm$); 
  and secondly a two-parameter model 
  in which $q(z)$ follows a step transition between two
  constant values, an early value
  $q_2$ to a late-time value $q_1$; also the transition redshift was
 artificially fixed at $z = 0.4$ or $0.6$, so this model set
 is quite restrictive and not very representative of plausible
  dark energy evolution.  
 Also, the title of \citet{hiz04} (R04) contains the phrase 
  {`Evidence for past deceleration...'}; that paper 
  is (as of 2014) the most-cited astrophysics paper published in 2004,
  and has thus been highly influential. 
 Specifically, R04 Section 4.1 considers a two-parameter model 
 $q(z) = q_0 + z (dq/dz)$ with constant $dq/dz$, and find that 
 $dq/dz$ is positive (implying past deceleration) 
 at above the 95\% confidence level. 
  Converting to $z_{acc} = -q_0 / (dq/dz)$, R04 derived 
 $z_{acc} = 0.46 \pm 0.13$.  However, we note that  
 there are several possible caveats in this result: firstly a constant
  $dq/dz$ model is somewhat unphysical since it leads to divergent 
  $q$ at large $z$; more realistic models like $\Lambda$CDM 
 have $dq/dz$ decreasing with $z$, so a linear $q(z)$ model 
  tends to underestimate $\zacc$;  a linear $q(a)$
 relation as in Section~\ref{sec:linqa}, \cite{cl08} and
 \cite{sca11} is probably more realistic. 
   Secondly, the choice of uniform priors in $q_0, dq/dz$ 
  leads to a prior density which is steeply rising towards small $\zacc$.
  If the true model is close to $\Lambda$CDM,  
   both of these effects may tend to pull the $\zacc$ estimate low. 
 Thirdly, as seen in \cite{cl08}, inclusion of more recent SNLS
  SN data also shifts the likelihood contours
   slightly towards smaller $dq/dz$; 
  most of their samples exhibit some non-decelerating regions inside
  the 95\% confidence contour.  
 
 From our discussions above, it appears that there has been a tendency to 
  overstate the strength of evidence for actual past 
  deceleration in SN data; 
  the results of R04, \cite{cl08}, \cite{mort-clark} and this paper 
  agree that there is reasonable evidence ($\sim 2 \sigma$) that
   $q(z)$ was less negative in the past than today, but the 
  GR-independent evidence for an actual zero-crossing
  (i.e. transition to deceleration) 
  is relatively weak, and sensitive to the choice of parametric form
  for $q(z)$.   
   Thus improved data is highly desirable to prove 
  past deceleration at high confidence.  
}

\section[]{A simple approximation for $D_L$}
\label{app:dlapp} 

Here, we note that we can also use the fitting function 
 equation~(\ref{eq:quadu}) to obtain a simple  
 `computer-free' approximation to $D_L(z)$ for 
 constant-$w$ $w$CDM models near the concordance model, 
 which is remarkably accurate up to $z < 1.7$. The procedure 
 goes as follows:   
\begin{enumerate} 
\item 
  For given $\Omm, w$, the standard  Friedmann equation gives the value of 
  $\zacc$ as in equation~(\ref{eq:zacc}),  hence $(1+\zacc)/E(\zacc)$ follows.  
\item 
 Given those two values above, we can then 
  readily solve for the pair $(b_1, b_2)$ in equation~(\ref{eq:quadu}) 
  which reproduce the same 
  position and value of the turning point in $(1+z)/E(z)$;  
  the result is $b_1 = 2 [(1+\zacc)/E(\zacc) - 1]/ \uacc$, 
  $b_2 = b_1 / 2 \uacc$  where $\uacc \equiv \ln(1+\zacc)$. 
\item
 Finally inserting the above constants $b_1, b_2$ 
  in equation~(\ref{eq:dlquad}) gives our
  simplified approximation for $D_L$. 
\end{enumerate} 
 Since this procedure matches only the turning point in 
 $(1+z)/E(z)$, as expected it results in slightly different
 values of $b_1, b_2$ compared to the previous
 case in \S~\ref{sec:quad} where we 
  numerically fitted $b_1,b_2$ to the 
 Friedmann $(1+z)/E(z)$ function over the full range $0 < u < 1$.  
 Thus we get a less accurate approximation to $D_L$, 
  but the accuracy still turns out surprisingly good 
 for this back-of-envelope level approximation.  For the case of the  
 concordance model $\Omm = 0.30$, $w = -1$, the recipe above 
  gives $b_1 = 0.5960$, $b_2 = 0.5804$, 
 hence the approximation becomes 
 \begin{equation} 
\label{eq:dlappa} 
  D_L(z) \simeq (c/H_0)(1+z) (u + 0.298 \, u^2 - 0.1935 \, u^3) \ ; 
 \end{equation}  
 comparing this to the quasi-exact numerical $D_L$ gives an rms error 
 of 0.14~percent and a worst-case error 0.32~percent
 across the range $0 < u < 1$  ($z < 1.72$).  
 The accuracy improves for $w > -1$ and
  degrades for $w < -1$, but remains $< 0.3$ percent rms 
 across the preferred ranges $0.27 < \Omm < 0.33$, $-1.2 < w < -0.8$. 
 The recipe above is substantially more accurate than a traditional
  third-order Taylor expansion in $z$, which rapidly becomes 
 poor at $z > 1$. 


\section[]{Relation between $\bmath{\lowercase{q}}$ 
 and jerk} 
\label{app:qjerk} 

Here, we provide a short argument why the simple fitting function 
 of equation~(\ref{eq:quadu}) works surprisingly well at $u < 1$.  
With the deceleration parameter $q$ defined as above, and the
 dimensionless jerk parameter $j$ defined by 
 \begin{equation} 
 j \equiv { d^3 a/dt^3 \over a H^3(a)} \ ,
\end{equation} 
 it is shown by e.g. \citet{bolotin} that 
\begin{eqnarray} 
  \frac{dq}{du} & = & j - q(2q+1) \nonumber \\ 
               & = & \frac{1}{8} + j - 2\left( q+\frac{1}{4} \right)^2 
\end{eqnarray} 
 (which is model-independent, assuming only that the derivatives exist). 
 In $\Lambda$CDM models, $j = +1$ independent of time (for negligible
  radiation  content);  this implies that $dq/du$ had a maximum when 
 $q = -1/4$, and was slowly varying between 7/8 and 9/8 
 over the period with $-0.60 < q < +0.10$, which corresponds to $z < 0.86$ and 
 $u < 0.62$ in our default model.  
 Also, differentiating equation~(\ref{eq:qu}) gives 
\begin{equation} 
 \frac{dq}{du} = - \frac{d^2}{du^2} \ln \left( {1+z \over E(z)} \right) \ ;
\end{equation} 
 thus a slowly-varying $dq/du$ leads to near-quadratic dependence for 
  $(1+z)/E(z)$ versus $u$.  
  We note that this 
 is partly coincidental for parameters near the concordance model, 
 since the present-day value $q_0 \simeq -0.55$ is near the end of 
 the timespan when $+0.1 > q > -0.6$. 

\bsp

\label{lastpage}

\end{document}